%% file: main.tex
\documentclass[twocolumn,epjc3]{svjour3}
\smartqed
\usepackage[utf8]{inputenc}
\usepackage{microtype}
\usepackage{amsmath}
\usepackage[english]{babel}
\usepackage{csquotes}
\usepackage[style=iso,showisoZ,useregional=false]{datetime2}
\usepackage{dirtytalk}
\usepackage{graphicx}\graphicspath{{./img/}}
\usepackage{nicefrac}
\usepackage{physics}
\usepackage{siunitx}
\usepackage{caption}
\usepackage{booktabs}
\usepackage[
    bibstyle=ieee,
    citestyle=numeric-comp,
    sorting=none,
    date=year,
    doi=true,
    isbn=false,
    eprint=true,
    url=true,
    maxnames=3, minnames=1,
]{biblatex}
\AtEveryBibitem{%
    \ifentrytype{article}{%
        \clearfield{title}
    }
}

\input{abbreviations}

\journalname{Eur. Phys. J. C}

\begin{document}

\title{Liquid argon light collection and veto modeling in GERDA Phase II}
\input{authors}
\date{}

\maketitle

\abstract{%
The ability to detect \acl{lar} scintillation light from within a densely packed \acl{hpge} detector array allowed the \acs{gerda} experiment to reach an exceptionally low background rate in the search for \acl{onbb} decay of \nuc{Ge}{76}.
Proper modeling of the light propagation throughout the experimental setup, from any origin in the \acl{lar} volume to its eventual detection by the novel light read-out system, provides insight into the rejection capability and is a necessary ingredient to obtain robust background predictions. 
In this paper, we present a model of the \acs{gerda} \acl{lar} veto, as obtained by Monte Carlo simulations and constrained by calibration data, and highlight its application for background decomposition.
}

\acresetall

\section{Introduction}\label{sec:introduction}

Provided with an array of germanium detectors, made from isotopically enriched \ac{hpge} material suspended in a clean \ac{lar} bath, the \ac{gerda} experiment set out to probe the neutrino's particle nature in a search for the \ac{onbb} decay of \nuc{Ge}{76}~\cite{Ackermann2013}.
The ability to detect scintillation light emerging from coincident energy depositions in the \ac{lar}, combined with \ac{psd} techniques~\cite{GERDA:2022ixh}, allowed to cut the background level of the second phase (\ptwo) to a record low~\cite{Agostini2018}. 
No signal was found, which translates into one of the most stringent lower limits on the half-life of the \ac{onbb} decay of \nuc{Ge}{76} at \SI{1.8e26}{\year} at \SI{90}{\percent} \acs{cl}~\cite{Agostini2020a}.

Based on dedicated Monte Carlo simulations of the scintillation light propagation, the model of the \ac{lar} veto rejection grants insight into the light collection from various regions of the highly heterogeneous setup. 
The full methodology and its first application are described in this document, which is structured as follows: 
\myref{sec:instrumentation} offers a brief description of the \ac{gerda} instrumentation, focusing on the \ptwo light read-out system. 
In \myref{sec:photon-detection-probabilities} the connection between photon detection probabilities and event rejection is made. 
\myref{sec:monte-carlo-implementation} summarizes the Monte Carlo implementation and the chosen optical properties, while \myref{sec:parameter-optimization} describes the tuning of the model parameters on calibration data.
In \myref{sec:probability-maps} photon detection probability maps are introduced, and in \myref{sec:background-decomposition} their application for background decomposition is highlighted.
In \myref{sec:conclusions} conclusions are drawn.

\section{Instrumentation}\label{sec:instrumentation}

The \ac{gerda} experimental site was the Hall~A of the INFN \ac{lngs} underground laboratory in central Italy.
Equipped with a large-scale shielding infrastructure \waitforit a \SI{64}{\cubic\meter} cryostat inside a \SI{590}{\cubic\meter} water tank \waitforit \ac{gerda} enclosed a low-background \ac{lar} environment, which from \DTMenglishmonthname{\DTMfetchmonth{startptwo}} \DTMfetchyear{startptwo} to \DTMenglishmonthname{\DTMfetchmonth{endptwop}} \DTMfetchyear{endptwop} gave home to the heart of \ptwo: 40, later 41, \ac{hpge} detectors in a 7-string array configuration, surrounded by a light read-out instrumentation. 
The veto design comprised two sub-systems: low-activity \acp{pmt}~\cite{Agostini2015c} and \ac{wls} fibers coupled to \acp{sipm}~\cite{Janicsko2016}.
The latter was upgraded in spring \DTMfetchyear{endptwo}.%\footnote{%
%The analysis presented in this manuscript concerns the initial, \ie pre-upgrade, fiber/\ac{sipm} installation, but is not limited to this application.}

\begin{figure}
    \centering
    \includegraphics{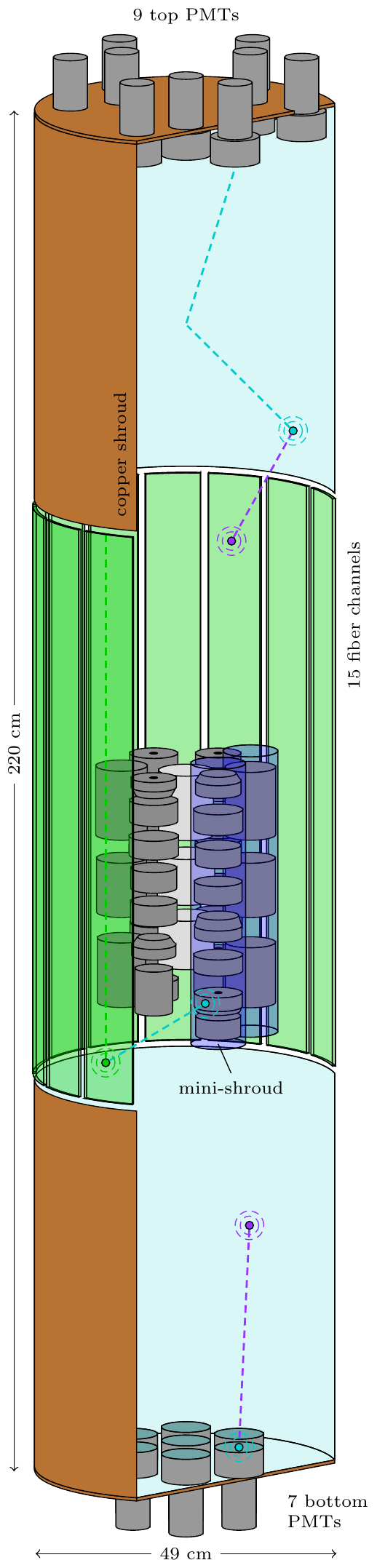}
    \caption{%
\acs*{lar} veto instrumentation concept. 
Transport of light signals towards the \acp{pmt} or \acp{sipm} relies on \acl{wls} processes in the \ac{tpb} layers or optical fibers. 
Several potential light paths are indicated. 
Support structure details, electronics as well as individual fibers are not drawn.%
    }\label{fig:lar-veto} 
\end{figure}

The \num{3}" Hamamatsu R11065-20 Mod \acp{pmt} were selected for their performance at \ac{lar} temperature and enhanced radiopurity~\cite{Acciarri:2011qx,Benson2018}.
Still, they contributed significantly to the background within the inner \ptwo setup and were thus placed at \SI{>1}{\meter} from any \ac{hpge} detector, giving \waitforit together with the limited cryostat entrance width \waitforit the \ac{lar} instrumentation its elongated cylindrical shape.
With space for support and calibration sources to enter, 3 off-center groups of 3 \acp{pmt} each were installed on top, whereas the bottom plate held 7 centrally mounted \acp{pmt}. 
Each \ac{pmt}'s entrance window was covered with \ac{tpb} embedded in polystyrene, in order to shift the incident \ac{vuv} scintillation light from \ac{lar} to a detectable wavelength. 
On the inside, the horizontal copper support plates were covered with a highly reflective \ac{tpb}-painted VM2000 multi-layer polymer, whereas lateral guidance of light towards the top/bottom was ensured by a \ac{tpb} dip-coated diffuse-reflecting \tetratex \ac{ptfe} foil~\cite{Baudis2015b}, stitched to the \SI{100}{\micro\meter}-thin copper shrouds.
A sketch of the setup is shown in \myref{fig:lar-veto}. 

With typical \acp{sipm} having a photo-sensitive area of \orderSI{1}{\square\centi\meter}, large-scale installations of \SI{>1}{\square\meter} photo coverage still represent a technological challenge~\cite{Kochanek2020}. 
Nonetheless, coupled to \ac{wls} fibers of \SI{<0.1}{\milli\becquerel\per \kilo\gram} activity in both \nuc{Th}{228} and \nuc{Ra}{226}, which serve as radio-pure light collectors~\cite{Janicsko2011}, the detection power of a single device is largely enhanced. 
The active chip size of the KETEK PM33100 \acp{sipm} is \noparseSI{3\times3}{\square\milli\meter}, they feature \SI{100}{\micro\meter} micro cell pitch and were purchased \say{in die}, \ie without packaging, allowing for a custom low-activity housing. 
Each of the \num{15} channels was comprised of \num{6} \acp{sipm}, connected in parallel on copper-laminated PTFE holders and cast into optical cement, amounting to a total active surface of \SI{8.1}{\square\centi\meter}.
The doubly-cladded BCF-91A fibers of square-shaped \noparseSI{1\times1}{\square\milli\meter} cross section, were covered with \ac{tpb} by evaporation, routed vertically to cover the central veto section, bent by \SI{180}{\degree} at the bottom and coupled to different \acp{sipm} on both top ends. 
Guidance of the individual fibers was ensured by micro-machined copper holders, attempting to keep them at a \SI{45}{\degree} rotation, facing their full \noparseSI{\sqrt{2}}{\milli\meter}-diagonal towards the center. 
%The total of $15\times6\times9 / 2 = 405$ fibers of about \SI{1.8}{\meter} length summed up to a total fiber length of about \SI{730}{\meter}.
The total length of the \num{405} fibers was about \SI{730}{\meter}.

Each of the \SI{40}{\centi\meter}-long \ac{hpge} strings was enclosed in a nylon \say{mini shroud}, transparent to visible light, covered on both sides with \ac{tpb}. It provided a mechanical barrier that limited the accumulation of \nuc{K}{42} ions \waitforit a progeny of cosmogenic \nuc{Ar}{42} \waitforit on the \ac{hpge} detector surfaces~\cite{Lubashevskiy2018}. 

The data acquisition of the entire array, based on SIS3301 Struck~\cite{Struck} FADCs, including the \ac{lar} veto photo sensors, was triggered once the signal of a single \ac{hpge} detector exceeded a pre-set online threshold. 
No independent trigger on the light read-out was implemented. 
The veto condition was evaluated offline, allowing for time-dependent channel-specific thresholds right above the respective noise pedestals and an anti-coincidence window that takes into account the characteristic scintillation emission timing as well as the \ac{hpge} detector signal formation dynamics.

\section{Photon detection probabilities}\label{sec:photon-detection-probabilities}

Upon interaction of ionizing radiation, ultra-pure \ac{lar} scintillates with a light yield of \SI{40}{photons\per\keV}~\cite{Doke2002}. 
There is an ongoing discussion whether this number could be smaller~\cite{Neumeier:2015lka}, but in any case, the actual light output is strongly reduced in the presence of trace contaminants and \apriori not precisely known for many experiments, including \ac{gerda}. 
An estimate based on the measured triplet lifetime of the argon excimer state of about \SI{1.0}{\micro\second}~\cite{Agostini2018}, limits the \ac{gerda} light yield to \SI{<71}{\percent} of the nominal pure-argon value, or \SI{<28}{photons\per\keV}.\footnote{%
This estimate only considers contaminant-induced non-radiative de-excitation to compete with the triplet decay and neglects impurities that effect the initial excimer production.}
%\footnote{%
%In first order only the long-lived triplet state is affected by contaminant-induced non-radiative de-excitation. 
%Electromagnetic interactions have a singlet-to-triplet ratio of \num{0.3}~\cite{Hitachi1983}. 
%The integral over the triplet decay for nominal \SI{1.6}{\micro\second}~\cite{Hitachi1983} \vs the quasi-reduced value of \SI{1.0}{\micro\second}~\cite{Agostini2018} results in a reduction of the light yield to $(0.3+1.0/1.6)/1.3=\SI{71}{\percent}$. 
%This estimate neglects impurities that effect the primary excimer production.}

Given a light yield $L'$ of this order, the number of primary \ac{vuv} photons produced in a typical \ac{gerda} background event can be enormous. 
Coincident energy depositions due to \be and \ga interactions in \ac{lar} (\eg from \nuc{Th}{228} or \nuc{U}{238} trace impurities) frequently reach \si{\MeV}-energies. 
The computational effort to track all \orderSI{10^{4}}{} optical photons represents a challenge, especially when considering the feedback between rejection power and required statistics \waitforit the larger the coincident energy release, the larger the suppression, the larger the statistics required to obtain a proper prediction of the \ac{hpge} spectrum after veto application. 
However, there is a workaround for this problem: the light propagation can be separated from the simulations that provide the energy depositions in the \ac{lar}.

The number of primary photons $n'$ generated from a single energy deposition $(E,\vec{x})$ in the \ac{lar}, follows a Poisson distribution $\mathcal{P}_{n'}(\lambda) = \lambda^{n'}e^{-\lambda}/n'!$ with expectation value $\lambda = E \cdot L'$.
Each of these photons has the opportunity to get detected with a photon detection probability \larmap, specific for interaction point $\vec{x}$. 
Accordingly, the number  of detected photons $n$, is the result of $n'$ Bernoulli trials, and stays Poisson distributed with expectation value $E \cdot L' \cdot \larmap$. 
Given a full event, with total coincident energy in the \ac{lar} distributed over several interaction points $(E_i,\vec{x_i})$, the \ac{pmf} $\lambda_s[n]$ for the total number of signal photons $n=\sum_i n_i$ reads
\begin{equation}\label{eq:veto}
 \lambda_s [n] = \mathcal{P}_n \Big( \sum_i E_i \cdot L' \cdot \xi(\vec{x_i}) \Big).
\end{equation}
As the convolution of several independent Poisson processes, it stays a Poisson distribution described by the sum of the expectation values.
Provided that \larmap is known, veto information can be provided on the basis of the underlying energy depositions $(E_i,\vec{x_i})$, and does not require optical simulations. 
It relies on the assumption that each set of photons, born from a particle's energy depositions $E_i$, solely depends on the primary light yield $L'$ and is emitted isotropically.\footnote{%
These assumptions break down in the case of particle-type dependent quenching or Cherenkov radiation emission. 
Both effects, as well as an explicit Fano factor, could be added to \myref{eq:veto} for future studies.
Recent studies suggest the absence of further energy-dependent quenching effects in \ac{lar}~\cite{Agnes:2018mvl}.}

Experiments using scintillation detectors traditionally quote the yield of detected \acp{pe} per unit of deposited energy, \ie the experimental light yield, in \eg \si{\pe\per\keV}. 
This number is only meaningful, when considering a homogeneous detector, with uniform response over most of its volume. 
By construction this is not the case for the \ac{gerda} \ac{lar} light read-out system, whose purpose is to detect light that emerges from within and around the optically dense \ac{hpge} detector array. 
With this in mind, good veto performance does not necessarily go hand-in-hand with maximum experimental light yield, especially when considering background that deposits energy in the \say{darkest} corners of the array, where the detection probabilities are minimal, possibly even zero. 
Hence, it is necessary to determine the full three-dimensional map of light detection probabilities \larmap, with special emphasis on the areas where little light is collected from.
This can only be done in a dedicated Monte Carlo study, that takes into account the full photon detection chain of the \ac{gerda} \ac{lar} instrumentation.

\subsection{A simple estimate}\label{sec:a-simple-estimate}

\begin{figure}
    \centering
    \includegraphics{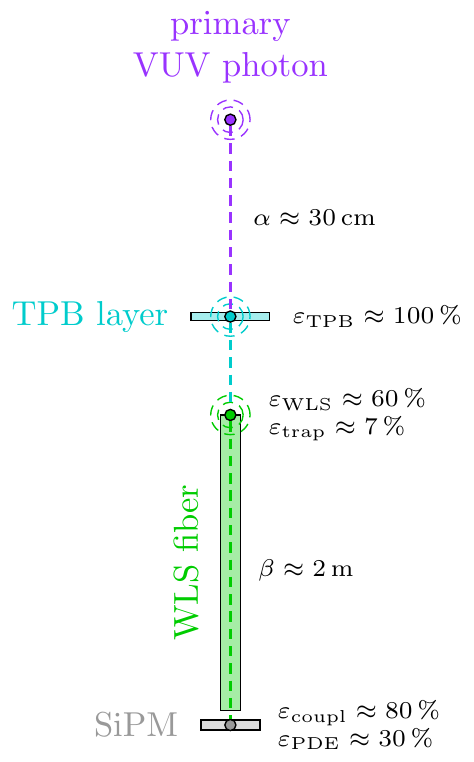}
    \caption{%
Simplified light collection chain. 
This one-dimensional representation depicts the main material properties that affect the light collection with the \ac{gerda} fiber-\acs*{sipm} instrumentation. 
The overall light collection efficiency for the primary \acs*{vuv} photon is of \orderSI{0.1}{\percent}.
In real life, effects like shadowing, reflections and optical coverage enter the game.
    }\label{fig:light-collection-chain} 
\end{figure}

Before running such simulations, it is worthwhile to evaluate the impact of the various steps a primary \ac{vuv} photon undergoes until its detection. 
\ac{gerda} uses a hybrid system consisting of \ac{tpb}-coated \ac{wls}-fibers with \ac{sipm}-readout and \ac{pmt}s to detect the \ac{lar} scintillation light that emerges from in and around the \ac{hpge} detector array. 
If we neglect most geometric effects, the photon detection probability can be broken down into factors that depend on basic properties of the materials and components involved. 
Given a primary photon that is emitted in the system \acs{lar}-\acs{tpb}-fiber-\acs{sipm}, as depicted in \myref{fig:light-collection-chain}, the probability $\xi$ for its detection, can be described by
\[
    \begin{aligned}
        \xi \;\propto\;
            &\overbrace{e^{-x/\alpha(\lambda)}}^{\text{\ac{lar}}}
            \;\times\;\varepsilon_{\text{TPB}}(\lambda)
            \;\times\; \overbrace{\varepsilon_{\text{WLS}}(\lambda)\;\varepsilon_{\text{trap}}\;e^{-y/\beta(\lambda)}}^{\text{fiber}} \\
        \times\; &\underbrace{\varepsilon_{\text{coupl}}\;\varepsilon_{\text{PDE}}(\lambda)}_{\text{SiPM}}\;,
    \end{aligned}
\]
Where $\lambda$ is the photon wavelength. First, the \ac{vuv} photon has to travel a certain distance $x$ in \ac{lar}, while risking to get absorbed in interactions with residual impurities. 
The absorption length $\alpha(\lambda=\SI{128}{nm})$ at \ac{lar} peak emission is on the order of tens of centimeters, depending on the argon purity~\cite{ArDM:2016jbw,Neumeier:2015lka,Barros2020}. 
The moment the \ac{vuv} photon reaches and gets absorbed in any \ac{tpb} layer, a blue photon with peak emission at \SI{420}{\nm} is re-emitted~\cite{Gehman2011}. 
The efficiency $\varepsilon_\text{TPB}$ for this process is close to \SI{100}{\percent}~\cite{Benson2018,Araujo:2021buv}.
Since the typical distance for a first encounter with a \ac{tpb}-coated surface is of similar order as the absorption length itself, about $1/e$ of the primary photons make it through this first part of the journey. 
Once a photon is shifted to blue, absorption in the \ac{lar} becomes negligible, as the absorption length for visible light exceeds the actual system size.
Hence, it does not necessarily matter, if the blue photon directly enters a fiber at this point or later. 
As soon as this is the case, the photon undergoes a second wavelength-shifting step and is shifted to green with peak emission at \SI{494}{\nm}~\cite{bcf91a}. 
The corresponding efficiency $\varepsilon_\text{WLS}$, \ie the overlap between the \ac{tpb} emission and fiber absorption spectrum, is about \SI{60}{\percent}. 
The green photon will stay trapped within the fiber with a trapping efficiency $\varepsilon_\text{trap}$ of about \SI{7}{\percent}~\cite{bcf91a} and arrive at its end after about half of its absorption length of $\beta\approx\SI{2}{\m}$, which adds another factor $1/\sqrt{e}$. 
The coupling efficiency $\varepsilon_\text{coupl}$ to successfully couple the photon into the \ac{sipm} is assumed to be $\SI{80}{\percent}$, whereas the \ac{pde} of being detected as a photo-electron signal is about \SI{30}{\percent} at the green fiber emission~\cite{pm33100}. 
Multiplication of all individual contributions results in an overall detection probability of not more than \SI{0.2}{\percent} and it can be anticipated that including geometric effects (\eg shadowing or optical coverage) the light collection will not exceed \SI{0.1}{\percent} for most regions of the \ac{gerda} \ac{lar} volume.
%\todo{check for number of map that can be anticipated here}

\section{Monte Carlo implementation}\label{sec:monte-carlo-implementation}

The \ac{gerda} instrumentation is implemented in the \acsu{geant}-based~\cite{Agostinelli2003,Allison2006,Allison2016} \acfu{mage} simulation framework~\cite{Boswell2011}. 
For what concerns the propagation of optical photons from typical background processes, most important are the geometries enclosed by the \ac{lar} veto instrumentation as well as the optical properties of the corresponding materials.

\subsection{Geometry}\label{subsec:geometry}

The \ac{hpge} detector array, including all auxiliary components, is implemented to the best available knowledge, but making reasonable approximations.
The reader may find detailed technical specifications such as dimensions and materials documented in \cite{Agostini2018}.
The simulated setup includes: individually sized and placed \ac{hpge} detectors in their silicon/copper mounts, \ac{tpb}-covered nylon mini-shrouds around each string, high-voltage and signal flat cables running from each detector to the front-end electronics, the front-end electronics themselves as well as copper structural components. 
Approximations are made when full degeneracy of events originating from the respective parts is expected, \eg the level of detail of the electronics boards is low and no detailed cable routing is implemented. 
Details are discussed in~\cite{Agostini2020}. 
As a consequence, shadowing effects that impact the optical photon propagation, but not the standard background studies, may not be captured perfectly. 

The \ac{pmt}s are implemented as cylinders, with a quartz entrance window and a photo-sensitive cathode.
They are placed at their respective \num{9}(\num{7}) positions in the top(bottom) copper plate, which to the inside is covered with a specular reflector that emulates VM2000. 
In contrast, the inside of the copper shrouds is lined with a diffuse \ac{ptfe} reflector that represents the \tetratex foil. 
All reflector surfaces, as well as the \ac{pmt} entrance windows, are covered with a wavelength-shifting \ac{tpb} layer.

The fiber shroud is modeled as $15\times6 = 90$ cylinder segments covering the central part of the veto volume. 
Every segment contains a core, two claddings and one thin \ac{tpb} layer, just as the real fibers. 
Their bottom ends have reflective surfaces attached, whereas optical photons reaching the upper ends are registered by photosensitive surfaces, each of them representing one \ac{sipm}. 
This differs from the real-world implementation, where the fibers are bent, up-routed and read-out on both ends. 
To avoid any misinterpretation of in-fiber correlations between channels that are connected to the same fibers, both the Monte Carlo and data signals are re-grouped to represent one channel per fiber module, resulting in a total of \num{9} fully independent channels. 
Due to sagging and uneven tensioning the optical coverage of the fibers was reduced in comparison to the maximum possible value of \SI{75}{\percent}.
In the simulation, a gap between the fiber segments parametrizes the coverage of the fiber shroud.
Analyzing photos of the mounted fiber modules the real coverage was estimated to be around \SI{50}{\percent}.
%\footnote{%
%Ideally all 810 fibers contribute their full \noparseSI{\sqrt{2}}{\milli\meter} lateral cross section to cover the $\pi \cdot \SI{470}{\milli\meter}$ circumference of the veto cylinder. 
%However, as not every fiber may follow a perfectly straight path in between the holders, the actual coverage can be significantly reduced and moreover non-uniform.}

\subsection{Optical properties}\label{subsec:optical-properties}

\begin{figure}
    \centering
    \resizebox{0.49\textwidth}{!}{\includegraphics{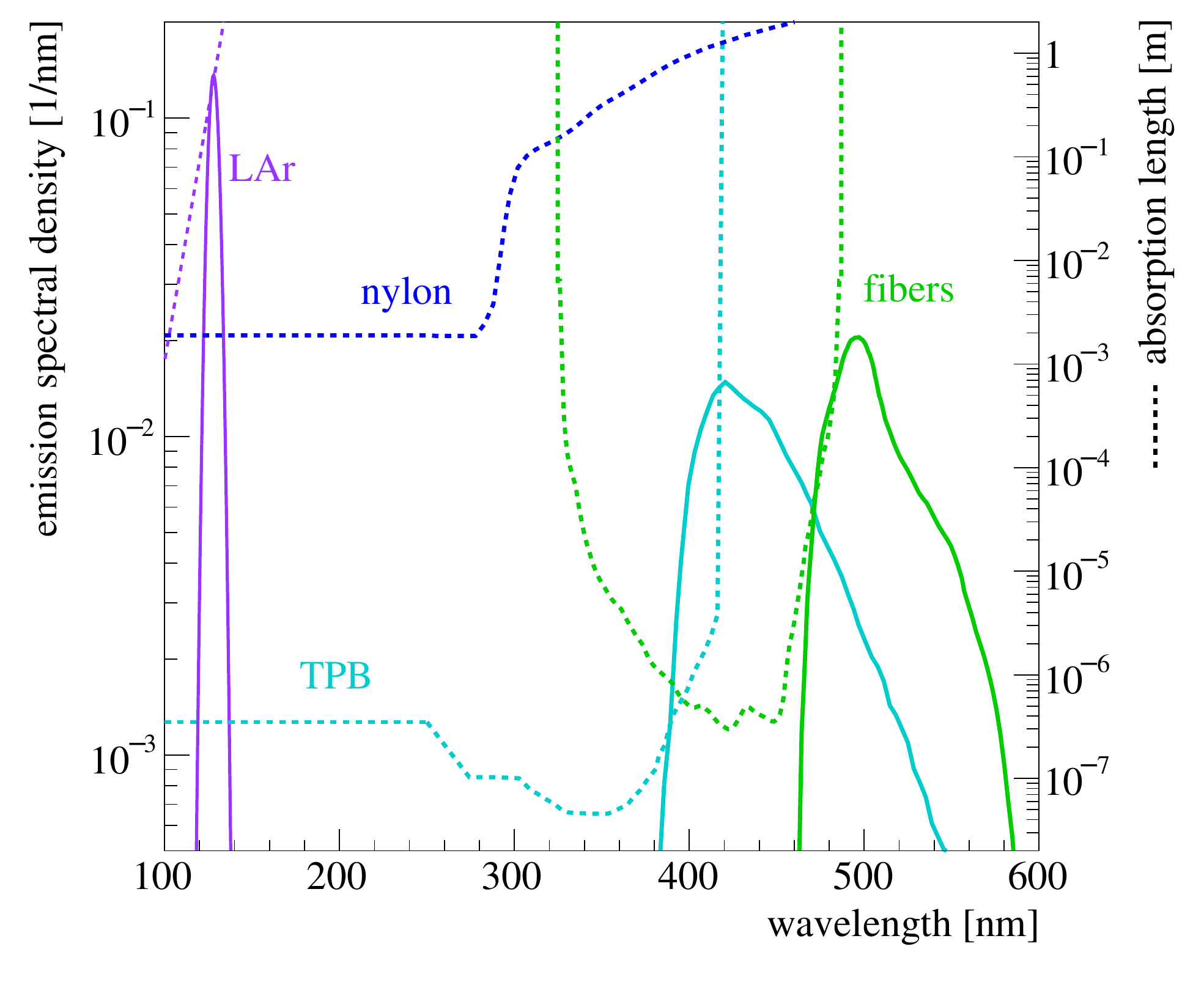}}
    \caption{%
Emission spectra (solid lines) and absorption length (dashed lines) of indicated materials.
The primary emission from the \ac{lar} follows a simple Gaussian distribution centered at \SI{128}{\nano\meter}. 
Its absorption length connects to larger wavelength with an \adhoc exponential scaling. 
Absorption and re-emission appears in \ac{tpb} and the polystyrene fiber material. 
Nylon is only transparent to larger wavelength.
    }\label{fig:wls-chain} 
\end{figure}

\myref{fig:wls-chain} compiles the relevant emission and absorption features implemented for the various materials. 
The emission of \ac{vuv} scintillation photons from the \ac{lar} follows a simple Gaussian distribution centered at \SI{128}{\nano\meter} with a standard deviation of \SI{2.9}{\nano\meter}. 
It neglects contributions at longer wavelength, which have orders of magnitude lower intensity for pure \ac{lar}~\cite{Heindl2010}. 
The refractive index of the \ac{lar} is implemented using the empirical Sellmeier formalism, with the coefficients obtained in~\cite{Bideau-Mehu1981}. 
Building on this, the wavelength-dependent Rayleigh scattering length is derived~\cite{Seidel2002}. 
It corresponds to about \SI{70}{\centi\meter} at \ac{lar} peak emission, which is shorter than recently suggested~\cite{Babicz2019}. 
Operation of the \acfu{llama} during the \ac{gerda} decommissioning point towards a \ac{vuv} attenuation length of about \SI{30}{\centi\meter}~\cite{Schwarz2021}. 
Accordingly, the absorption length was set to $1/(1/30-1/70) \approx \SI{55}{\cm}$. 
The absorption length is modeled over the full wavelength range extending it from \SI{128}{\nano\meter} with an \adhoc exponential function.
The primary \ac{vuv} scintillation yield $L'$ is considered a free parameter and by default set to \SI{28}{photons\per\keV}.
A dependence of the photon yield on the incident particle, \ie quenching, as well as characteristic singlet and triplet timing are implemented. 
The \ac{tpb} absorption length is taken from~\cite{Benson2018}, the emission spectrum from~\cite{Gehman2011}. 
Individual emission spectra, where available, are implemented for \ac{tpb} on nylon~\cite{Lubashevskiy2018}, VM2000~\cite{Francini2013} as well as \tetratex~\cite{Baudis2015b}. 
The absorption length of nylon is taken from~\cite{Agostini2018a}. 
Absorption and emission of the fiber material use the data presented in~\cite{bcf91a}, normalized to measurements at \SI{400}{\nano\meter}. 
The quantum efficiency of the \acp{pmt}~\cite{r11065-20} and \ac{pde} of the \acp{sipm}~\cite{pm33100} have been extracted from the product data sheets provided by the vendors. 
The reflectivities of germanium, copper, silicon and \ac{ptfe} above \SI{280}{nm} are taken from~\cite{wegmann}, whereas their values at \ac{vuv} wavelength are largely based on assumptions. 
The reflectivity of VM2000 is taken from~\cite{Francini2013}, the one of \tetratex from~\cite{Janecek2012}.
The exact optical property values implemented in the simulation have been reported in~\cite{pertoldi}.

\subsection{Uncertainties}

\Apriori, the bare simulations are not expected to reproduce the data.
Details like partially inactive \ac{sipm} arrays, coating non-uniformities and shadowing by real-life cable management are not captured by the Monte Carlo implementation. 
Similarly, input parameters measured under conditions differing from those in \ac{gerda}, \eg at room temperature or different wavelength, pose additional uncertainty.

Back to \myref{eq:veto} and photon detection probabilities \larmap: as already the number of primary \ac{vuv} photons is uncertain, any linear effect, constant across the \ac{lar} volume $\vec{x}$, is degenerate with the primary light yield $L'$ and thus only the product $L' \cdot \larmap$ can be constrained by data-Monte Carlo comparison. 
It follows that, if a primary light yield of $L' = \SI{28}{photons\per\keV}$ is assumed, its true value is fully absorbed in a global scaling of the efficiencies $\varepsilon_i$, individually to each light detection channel $i$. 
The set $\varepsilon_i$ may further absorb any other global effect, \eg an inaccurate \ac{tpb} quantum efficiency, as well as any local channel-specific feature, \eg varying photon detection efficiencies of the photo sensors. 
Given the large set of potential uncertainties, $\varepsilon_i$ are treated unconstrained and may take any value between zero and unity.
%Input parameters whose impact is non-linear across $\vec{x}$ and hence not described by simple scaling, are in principle accessible through statistical inference. 
%In practice, given the high degree of correlation between parameters that makes the estimation challenging, it is much more advisable to determine optical properties of interest through independent and dedicated measurement setups.

\section{Parameter optimization}\label{sec:parameter-optimization}

To obtain a predictive model of the performance of the \ac{lar} veto system, residual degrees of freedom must be removed.
In the following, we describe the methodology employed to statistically infer the value of the efficiencies $\varepsilon_i$ by comparing simulated to experimental data.
The full evidence of the model parameters is contained in a likelihood function, which has been maximized for special calibration data.

Given a class of events, the \ac{pmf} $\Lambda[n]$ that describes the number of photons $n$ detected by some \ac{lar} veto channel is the convolution of two contributions:
\begin{equation}\label{eq:pmf-convolution}
    \Lambda[n] = \Lambda_s[n] * \Lambda_b[n]\;.
\end{equation}
It is a simultaneous measurement of light from true coincidences $\Lambda_s[n]$ that accompany the corresponding \ac{hpge} energy deposition as well as random coincidences $\Lambda_b[n]$ largely produced by spectator decays such as \eg \nuc{Ar}{39} in the \ac{lar}.\footnote{%
It is not the pure Poisson distribution of \myref{eq:veto}, as it arises from all different realizations of coincident energy deposition in the \ac{lar} as $\Lambda[n] = \langle \lambda[n] \rangle$.}  
%Here $\lambda[n]$ is not described by a Poisson distribution in \myref{eq:veto} anymore since it statistically emerges from different realizations of energy depositions in the \ac{lar}. 
While $\Lambda_s[n]$ may be provided from simulations, randomly triggered events allow an evaluation of $\Lambda_b[n]$ from data.
However, as the measured signal amplitudes suffer non-linear effects, \eg afterpulsing and optical crosstalk, that are themselves under study and at present not implemented in the simulation, no direct \ac{pmf} comparison is possible and instead the binary projection of \myref{eq:pmf-convolution} is used.
In the binary \say{light/no-light} projection, where $\overline{\Lambda}=\Lambda[0]$ corresponds to no light, and $\Lambda$ to a positive light detection, the \ac{pmf} breaks down to a single expectation value, given by
\begin{equation}\label{eq:detection-probability}
    \begin{aligned} 
        \Lambda &= \Lambda_s \cdot \overline{\Lambda}_b + \overline{\Lambda}_s \cdot \Lambda_b + \Lambda_s \cdot \Lambda_b = \Lambda_s \lor \Lambda_b \\
        \overline{\Lambda} & = 1 - \Lambda = \overline{\Lambda}_s \cdot \overline{\Lambda}_b\;.
    \end{aligned}
\end{equation}
A positive light detection is either truly coincident without random contribution, fully random or a simultaneous detection of both. 
It is complementary to no detection, neither as true nor as random coincidence. 
$\Lambda_s$($\Lambda_b$) is the detection probability for true(random) coincidences and $\overline{\Lambda}_s = \overline{\Lambda} / \overline{\Lambda}_b$ quantifies the true survival probability of the underlying class of events, corrected for random coincidences.

\begin{figure}
    \centering
    \resizebox{0.49\textwidth}{!}{\includegraphics{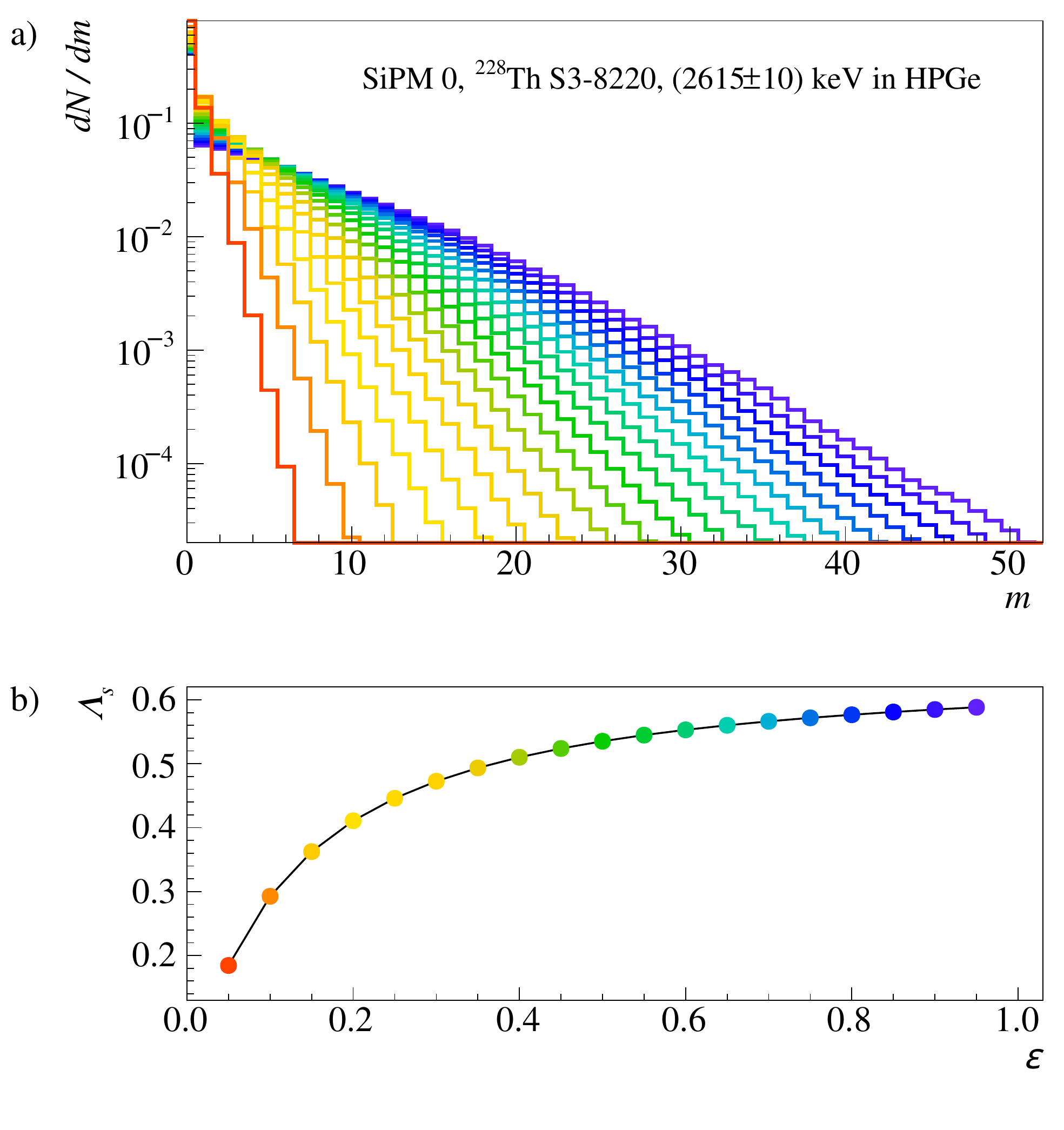}}
    \caption{%
Binomial repopulation. 
a) The \ac{pmf} $\Lambda_s[m](\varepsilon)$ defined in \myref{eq:binomial-repop} can be obtained for any value of the detection efficiency $\varepsilon$ from the unaltered simulation output.
   The example shows the \ac{pmf} for a specific \ac{sipm} channel as obtained for \nuc{Th}{228} decays in a calibration source at the top of the array (see also \myref{tab:pca-data}) depositing \SI{2615\pm10}{\keV} in the \ac{hpge} detectors.
   The result is plotted for a selection of efficiency values, reported in the bottom panel.
b) The panel shows the non-linear dependence of the light detection probability $\Lambda_s$ (defined in \myref{eq:det-prob}).
   The color coding relates data points with corresponding distributions in the top panel. 
    }\label{fig:binomial-repopulation} 
\end{figure}

Even though the data is reduced to binary information, the simulated \ac{pmf} $\Lambda_s$ allows the additional detection efficiency $\varepsilon$ to be folded into the Monte Carlo expectation. 
Given a count of $n$ photons in the bare simulation, an effective detection of $m<n$ photons can be represented as a sequence of Bernoulli trials with probability $\varepsilon$. 
The \ac{pmf} $\Lambda_s[m](\varepsilon)$ is the result of binomial re-population throughout all $n \geq m$: 
\begin{equation}\label{eq:binomial-repop}
    \Lambda_s[m](\varepsilon) = \sum_{n \geq m} \Lambda_s[n] \binom{n}{m} \varepsilon^m (1-\varepsilon)^{n-m}.
\end{equation}
This technique avoids re-simulation for different values of $\varepsilon$. 
\myref{fig:binomial-repopulation} depicts Monte Carlo spectra processed for different efficiencies. 
Back in binary space, the detection probability $\Lambda_s(\varepsilon)$, \ie the chance to see one photon or more as true coincidence, is
\begin{equation}\label{eq:det-prob}
    \Lambda_s(\varepsilon)
    = 1 - \overline{\Lambda}_s(\varepsilon)
    = 1 - \sum_{n} \Lambda_s[n] (1-\varepsilon)^{n} \;.
\end{equation}
It is the inverse of no detection, \ie the population of the \say{zero bin} $\Lambda_s[0](\varepsilon)$ in \myref{eq:binomial-repop} and allows uncertainties on the bare frequencies $\Lambda_s[n]$ to be propagated into $\Delta\Lambda_s(\varepsilon)$.\footnote{%
Defined as \(
  [\Delta\Lambda_s(\varepsilon)]^2 =
    \sum_n \{
      [(1-\varepsilon)^n - \overline{\Lambda}_s(\varepsilon)]
      {\Delta N_n}
      / N_\text{tot}
    \}^2
\), where $\Delta N_n$ is the uncertainty of the unaltered $n$-photon observations in a total of $N_\text{tot}$ Monte Carlo events.}

The likelihood for the observation of light in $N$ Monte Carlo events out of $N_\text{tot}$ simulated in total, given the aforementioned expectation value $\Lambda(\varepsilon)=\Lambda_s(\varepsilon) \lor \Lambda_b$, is described by a binomial distribution $\mathcal{B}^N_{N_\text{tot}}(\Lambda) = \binom{N_\text{tot}}{N} \Lambda^{N} (1-\Lambda)^{N_\text{tot}-N}$. 
Maximizing its value allows to infer on $\varepsilon$, whereas taking into account the limited statistics of the random coincidence dataset, with $M$ light detections over $M_\text{tot}$ random events, makes it a combined fit of both the data and the random coincidence sample. 
This combined likelihood reads:
\begin{equation}\label{eq:1d-likelihood}
    \begin{aligned}
        \mathcal{L}(\varepsilon;\sigma) = \;
            &\mathcal{B}_{N_\text{tot}}^{N}\big((\Lambda_s(\varepsilon)+\sigma\cdot\Delta\Lambda_s(\varepsilon))\lor\Lambda_b\big) \times \\
            &\mathcal{B}_{M_\text{tot}}^{M}\big(\Lambda_b\big) \times \hat{\mathcal{G}}\big(\sigma\big) \;.
    \end{aligned}
\end{equation}
The signal expectation is given flexibility according to its uncertainty $\Delta\Lambda_s$ using a Gaussian pull term $\hat{\mathcal{G}}(\sigma) = e^{-\sigma^2/2}$, which accounts for limited simulation statistics and additional systematics.
\myref{eq:1d-likelihood} has zero degrees of freedom and hence model discrimination can only be obtained through a combination of multiple datasets, \ie calibration source positions, or by exploiting channel event correlations. 
While the former sounds trivial, \eg an absorption length can be estimated from  measurements at different distance, the latter requires explanation.
Let's imagine two photosensors $i \in \{A,B\}$, each probing the \ac{lar} with a certain photon detection probability $\xi_i(\vec{x})$. 
Considering an energy deposition $(E,\vec{x})$, the probability to see light in both channels depends on $\xi_A(\vec{x}) \cdot \xi_B(\vec{x})$, while events triggering only channel $A$ test $\xi_A(\vec{x}) \cdot (1-\xi_B(\vec{x}))$.
Both of them probe distinct regions of the \ac{lar} volume.
The probability to see no light from a certain volume of the \ac{lar}, \ie the corresponding survival probability, depends on $\prod_i (1-\xi_i(\vec{x}))$.

Given the full set of veto channels $S$ of size $n$, each event will come as a certain subset, \ie pattern, $P \subseteq S$ of triggered channels.
The total number of possible patterns is $2^n$, where each of them comes with its own unique expectation value derived from signal as well as random coincidences. 
A pattern's signal expectation $\Lambda_s(\vec{\varepsilon})$ can be evaluated much like \myref{eq:binomial-repop}, however starting from an $n$-dimensional hyper-spectrum evaluated for the full vector of efficiencies $\vec{\varepsilon}$. 
When folding in the random coincidences, it has to be considered that a certain pattern $P_s=\{A,B\}$ may be elevated to \eg $P=\{A,B,C\}$ by random coincidences of the form $P_b=\{A,C\}$ or similar. 
Each pattern occurrence expectation value is hence a sum over all possible generator combinations $G=\{P_s,P_b\}$, that result in $P_s \lor P_b = P$. 
The full likelihood reads
\begin{equation}\label{eq:pattern-likelihood}
    \begin{aligned}
        \mathcal{L}(\vec{\varepsilon};\sigma) = \;
            &\prod_{P} \mathcal{B}_{N_{tot}}^{N}\big(\sum_{G}(\Lambda_s(\vec{\varepsilon})+\sigma\cdot\Delta\Lambda_s(\vec{\varepsilon}))\cdot\Lambda_b\big) \times\\
            &\prod_{G} \mathcal{B}_{M_{tot}}^{M}\big(\Lambda_b\big) \times \hat{\mathcal{G}}\big(\sigma\big) \;.
    \end{aligned}
\end{equation}
The number of degrees of freedom is $(2^n-n-1)$, where $n$ is the number of channels. 
Several datasets may be combined as the product of the individual likelihoods. 
\myref{eq:pattern-likelihood} implicitly includes correlations between photosensors, which avoids a potential overestimation of the overall veto efficiency that could arise from unaccounted systematics. 
Given a large set of channels the number of possible patterns may be immense, but can be truncated by \eg neglecting events with detection pattern higher than a certain multiplicity. 
%This cutoff can be employed to effectively reduce the computational complexity of $\mathcal{L}(\vec{\varepsilon};\sigma)$ with large datasets.

\subsection{Application}\label{subsec:application}

\begin{table}
    \centering
    \caption{%
Calibration data taken with a \SI{<2}{\kilo\becquerel} \nuc{Th}{228} source placed at different heights.
The reported position corresponds to the absolute distance moved from the parking position on top of the experiment.
The upper-most and lower-most \ac{hpge} detectors are situated at about \SI[parse-numbers = false]{8180}{\mm} and \SI[parse-numbers = false]{8560}{\mm} respectively.
    }\label{tab:pca-data}
    \begin{tabular}{ccc} 
        \toprule
        position     & live time      & random       \\
        {[\si{\mm}]} & {[\si{\hour}]} & coincidences \\
        \midrule
        \num{8220}   & \num{6.4}      &  \SI{ 7.5( 6)}{\percent} \\
        \num{8405}   & \num{4.3}      &  \SI{ 7.2(10)}{\percent} \\
        \num{8570}   & \num{3.6}      &  \SI{10.2(14)}{\percent} \\
        \bottomrule
    \end{tabular}
\end{table}

In order to constrain the aforementioned effective channel efficiencies $\varepsilon_i$, dedicated data taken under clear and reproducible conditions was needed. 
As the \ac{gerda} background data is a composition of various contributions and itself under study, only the peculiar conditions of a calibration run, where the energy depositions originate from a well-characterized source, allow for such studies. 
However, usual calibrations were taken with the purpose to guarantee a properly defined energy scale of all \ac{hpge} detectors and were performed with three \nuc{Th}{228} calibration sources of \orderSI{10}{\kilo\becquerel} activity each. 
The resulting rate in the \ac{lar} was far too high to study the veto response and was hence not even recorded during those calibrations. 
For this reason, special calibration data using one of the former \pone \nuc{Th}{228} sources with an activity of \SI{<2}{\kilo\becquerel} was taken in \DTMenglishmonthname{\DTMfetchmonth{pca68}} \DTMfetchyear{pca68}.
The source was moved to 3 different vertical positions.
The characteristics of the datasets are compiled in \myref{tab:pca-data}. 
To avoid \be particles contributing to the coincident light production and enable a clean \ga-only signature, an additional \noparseSI{3}{\mm} copper housing was placed around the source container.\footnote{%
The suppression achieved in this configuration is not representative for backgrounds detected during physics data taking, which mostly originate from thin low-mass structural components, where \be particles contribute to the energy depositions in the \ac{lar}.}
Each configuration was simulated with \num{e8} primary decays in the source volume.

\begin{figure}
    \centering
    \resizebox{0.49\textwidth}{!}{%
        \includegraphics{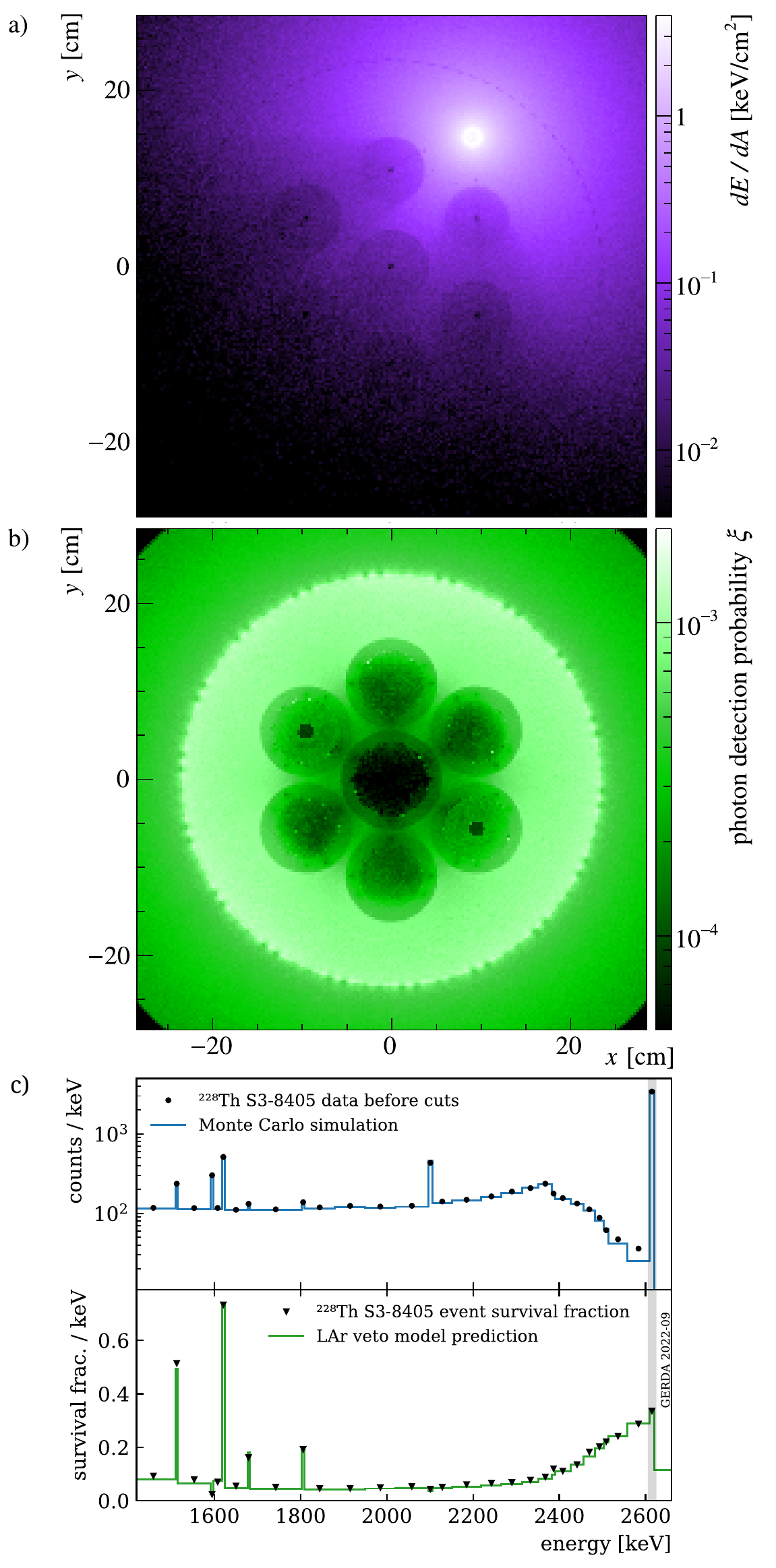}
    }
    \caption{%
Data/Monte Carlo comparison. 
a) Projected distribution of the \ac{lar} energy depositions for simulated \nuc{Tl}{208} \SI{2615}{\keV} \ac{fep}-events from a calibration source at position \SI{8405}{\mm}.
   Darker circles correspond to the volume occupied by the \ac{hpge} detectors.
b) Photon detection probability $\xi$ in the same region.
c) Top panel: the energy spectrum of the \nuc{Th}{228} data corresponding to figure \emph{a)}, before the \ac{lar} veto cut, compared to the Monte Carlo prediction. The \emph{pdf} is normalized to reproduce the total count rate in data. Despite small shape discrepancies, the predicted \ac{lar} veto survival probability (bottom panel) matches the data over a wide range of energies, even far from the model optimization energy region (gray band). A variable binning is adopted for visualization purposes.
}\label{fig:pcalib} 
\end{figure}

The maximum likelihood analysis was performed on \nuc{Tl}{208} \ac{fep} events with an energy deposit of  \SI{2615(10)}{\keV} in a single \ac{hpge} detector.
As no direct \be transitions to the ground nor first excited state of the \nuc{Pb}{208} daughter nucleus are allowed, a minimum of \SI{3.2}{\MeV} is released in \ga's, which almost always includes a transition to the intermediate \SI{2615}{\keV} state.
Selecting full absorption events of the corresponding \ga, results in an event sample virtually independent on \ac{hpge} detector details \waitforit the \ac{hpge} array is solely used to tag the \nuc{Tl}{208} transition.
The coincident energy depositions in the \ac{lar} originate to a large extent from coincident \ga's of \SI{583}{\keV} or more, and only marginally from \emph{Bremsstrahlung}.
\myref{fig:pcalib} a) shows their energy distribution in the \ac{lar}.
The random coincidence samples were obtained by test pulse injection at \SI{50}{\milli\hertz} as well as an early (\SI{-20}{\micro\second}) evaluation of the veto condition.
The random coincidence appearance is unique to each source configuration as the energy depositions from independent decays in the source contribute.

To reduce the dimensionality of \myref{eq:pattern-likelihood}, the top/bottom \ac{pmt} channels were regrouped to represent one single large top/bottom \ac{pmt}, whereas the generally less uniform fiber/\ac{sipm} channels were kept separate.
Accordingly, the likelihood had to be evaluated in $2 + 9 = 11$ dimensions. 
The pattern-space was truncated so that only channel combinations present in data had to be calculated.
The extracted channel efficiencies are \SI{13}{\percent} for the top \acp{pmt}, \SI{29}{\percent} for the bottom \acp{pmt} and reach from \SIrange[]{21}{37}{\percent} for the \ac{sipm} channels, with uncertainties of about \SI{\pm1}{\percent}.
Given an additional systematic uncertainty of \SI{20}{\percent} on the observation of the various veto patterns (\ie $\sigma = 0.2$ in \myref{eq:pattern-likelihood}), the p-value amounts to 0.2. 
The differences in the \ac{sipm} efficiencies match the expectation for problematic channels with potentially broken chips.
The reduced value for the top \acp{pmt} was anticipated, as additional shadowing effects from cabling are only present at the top.
%Assuming that all other efficiencies are correct, one could argue that the light yield of the \ac{gerda} \ac{lar} should be at around \SI{10}{photons\per\keV}, taking the originally assumed \SI{28}{photons\per\keV} and the maximum obtained channel efficiency.

\section{Probability maps}\label{sec:probability-maps}

To finally evaluate the three-dimensional photon detection probability \larmap, a dedicated simulation of \ac{vuv} photons sampled uniformly over the \ac{lar} volume around the \ac{hpge} array was performed.
Positive light detections with any photosensor $i$ were determined taking into account the efficiencies $\varepsilon_i$ as obtained in \myref{subsec:application}.
For convenience, \larmap is stored as a discrete map, partitioned into cubic \say{voxels} of size $3 \times 3 \times 3$\;\si{\cubic\mm}. 
This size matches the characteristic scale of the probability map gradients expected in \ac{gerda}. 
Hence, the detection probability $\xi_k$ associated with voxel $k$ corresponds to the ratio between positive light detections and total scintillation photons generated in the voxel volume. 
\myref{fig:pcalib}~b) shows a projection of this object.
As outlined in \myref{sec:photon-detection-probabilities} it allows to determine the expected number of signal photons and the corresponding event rejection probability, solely based on energy depositions in the \ac{lar}.
Outside the densely packed array \orderSI{0.1}{\percent}-level values are reached. 
\myref{fig:pcalib}~c) compares the energy distribution, before and after the \ac{lar} veto cut, of the \nuc{Th}{228} source data with the model prediction.
Small discrepancies are expected from geometry inaccuracies and the modeling of charge collection at the \ac{hpge} surface. However, the event suppression (shown in the bottom panel) is reproduced over a wide range of energies far off the \nuc{Tl}{208} \ac{fep}-events used for model optimization. 
As expected, the rejection power is reduced for single-\ga \acp{fep} without significant coincidences in the decay scheme, \eg for \nuc{Bi}{212} at \SI{1621}{\keV}, and enhanced for the \ac{dep} at \SI{1593}{\keV}, where two \SI{511}{\keV} light quanta leave the \ac{hpge} detectors.
%one of the low-activity calibration datasets before and after \ac{lar} veto application, comparing data and Monte Carlo, whereas the latter was evaluated using the photon detection probability map to draw a boolean veto classifier.
%Taking into account the dataset-specfic random coincidence expectation from \myref{tab:pca-data}, they agree over a wide range of energies, that exceeds the \ac{fep} region used for parameter optimization.

\subsection{Distortion studies}\label{sec:distortion-studies}

\begin{figure*}
    \centering
    \resizebox{0.99\textwidth}{!}{\includegraphics{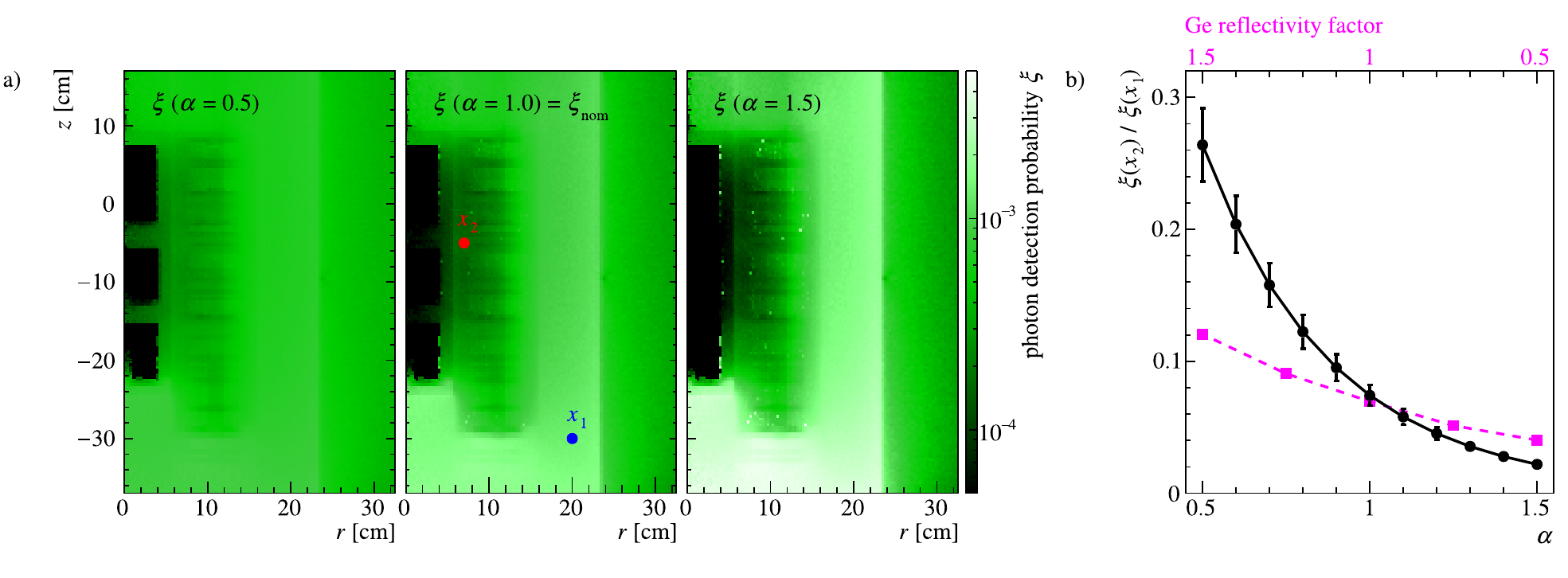}}
    \caption{%
Modification of the photon detection probability \larmap through analytical power-law distortions defined in \myref{eq:power-law}. 
a) Inhomogeneities that are present in the nominal map are amplified with increasing $\alpha$, leading to a more homogeneous ($\alpha<1$, \ie less color contrast) or less homogeneous ($\alpha>1$, \ie more color contrast) response.
b) The probability ratio from two sample points $x_1$ and $x_2$, outside and within the array, highlights this modification (black data points). 
The comparison of an altered germanium reflectivity (magenta data points) shows how the distortions conservatively exceed $\pm\SI{50}{\percent}$ on the reflectivity.
    }\label{fig:distortions}
\end{figure*}

As anticipated in \myref{sec:monte-carlo-implementation}, many uncertainties affect the simulation of the \ac{lar} scintillation in the \ac{gerda} setup. 
The channel efficiencies $\vec{\varepsilon}$ extracted from calibration data, as described in \myref{sec:parameter-optimization}, absorb systematic biases that scale the detection probability by global factors (\eg the \ac{lar} scintillation yield and the \ac{tpb} quantum efficiency), but cannot cure local uncertainties (\eg due to incorrect germanium reflectivity or fiber shroud coverage).
A heuristic approach has been formulated to estimate the impact of such simulation uncertainties on the LAr veto model. 
The distortion of \larmap induced by varying input parameters can be conservatively parametrized by means of an analytical transformation T: 
\[
    \larmap \mapsto \xi^\prime(\vec{x}). 
\]
As an additional constraint, the transformed $\xi^\prime(\vec{x})$ must still reproduce the calibration data presented in \myref{sec:parameter-optimization}, with which the original \larmap was optimized. 
As a consequence, the \ac{lar} volumes probed by calibration data (see \myref{fig:pcalib}) act as a fixed point of the transformation $T$, letting the detection probability deviate from its nominal value in all the other regions of the setup. 
$T$ may take various analytical forms, depending on the desired type of induced local distortions. 
The adoption of this procedure overcomes the difficulty of studying the dependence of \larmap on numerous optical parameters by performing several computationally expensive simulations of the scintillation light propagation.

In the context of this work, we shall focus on transformations that make \larmap less or more homogeneous, \ie that make \say{dark} areas (e.g.~the \ac{hpge} array) \say{darker} compared to areas with high detection probability, or \emph{vice versa}. 
A simple transformation that meets this requirement is the following power-law scaling:
\begin{equation}\label{eq:power-law}
  \larmap \mapsto N \cdot \larmap^\alpha \;,
\end{equation}
where $\alpha$ is a real coefficient controlling the magnitude of the distortion and $N$ is a normalization constant adjusted to reproduce the event suppression observed in calibration data. 
The action of the transformation in \myref{eq:power-law} on the detection probability is depicted in \myref{fig:distortions}.
In the same figure, the size of power-law distortions is compared to that induced by uncertainties on the \ac{hpge} detectors' reflectivity in the \ac{vuv} region.
The impact of the latter has been evaluated by scaling its value by $\pm\SI{50}{\percent}$ in dedicated optical simulations.
The power-law distortions significantly exceed the effect of a potential reflectivity bias and can therefore be used as a conservative estimate of the uncertainty on the light collection probability.

\section{Background decomposition}\label{sec:background-decomposition}

As an application of the light collection and veto model, we shall now present the results of the background decomposition of the \ac{hpge} energy spectrum recorded during physics data taking after the application of the \ac{lar} veto cut. 
This background model serves as a fundamental input for various physics analyses whose sensitivity is enhanced with the \ac{lar} veto background reduction.

Previous work~\cite{Agostini2020} has proven successful in describing the \ac{gerda} data in terms of background components, but before \ac{lar} veto and \ac{psd} cuts, referred to as \say{analysis} cuts. 
\Acp{pdf} for various background sources have been produced with dedicated simulations of radioactive decays in the inner \ac{gerda} setup. 
A linear combination of these \acp{pdf} has been consequently fit on the first \SI{60.2}{\kg\year} data from \ac{gerda} \ptwo in order to infer on the various contributions to the total background energy spectrum. 
Since the data was considered before the application of the \ac{lar} veto cut, the propagation of scintillation photons has been disabled in these simulations. 
The developed \ac{lar} detector model allows to incorporate this missing information into the existing simulations and compute background expectations after the \ac{lar} veto cut.
\Acp{pdf} for a representative selection of signal and background sources in the \ac{gerda} \ptwo setup are reported in \myref{fig:pdfs}.

\begin{figure*}
    \centering
    \resizebox{0.99\textwidth}{!}{\includegraphics{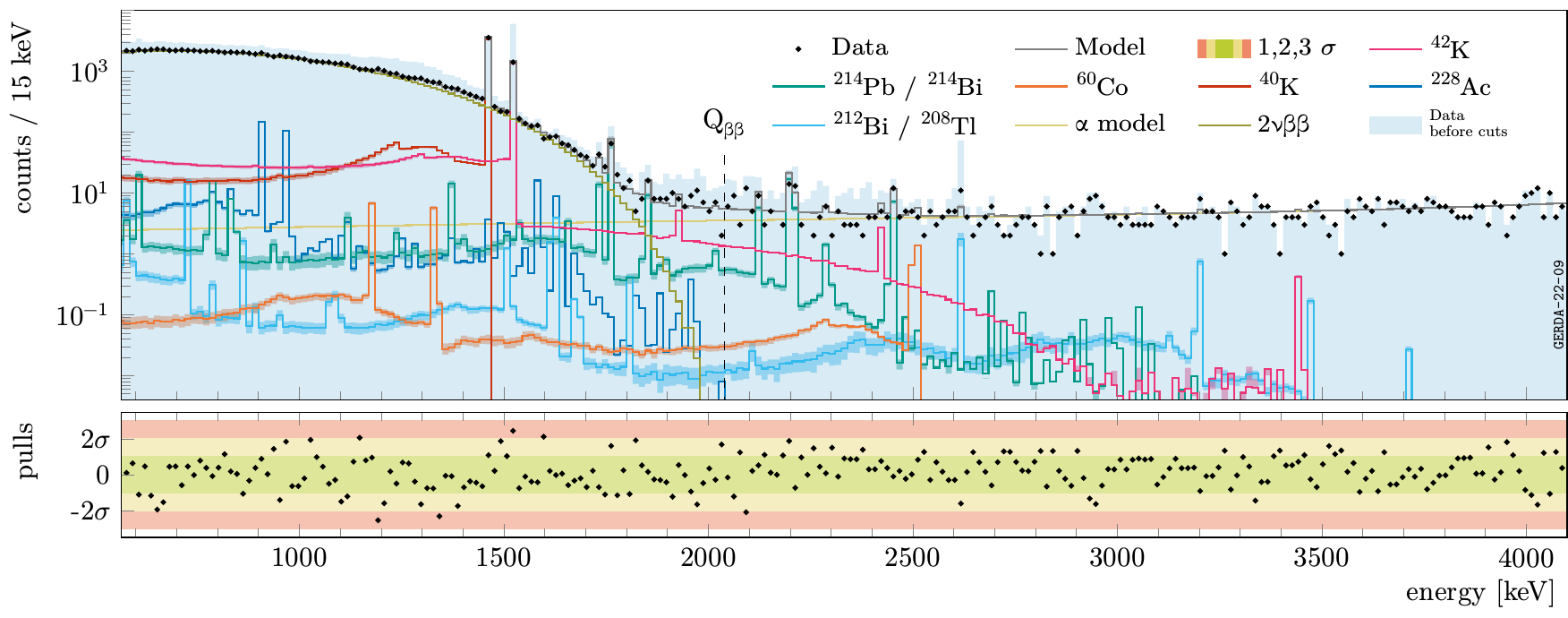}}
    \caption{%
Background decomposition of the first \SI{61.4}{\kg\year} of data from \ac{gerda} \ptwo surviving the \ac{lar} veto cut (black dots).
The veto model is applied to the existing background \acp{pdf} before the cut~\cite{Agostini2020} folding in the probability map \larmap.
Data before the cut is shown as a light blue filled histogram.
Shaded bands constructed with the maximally distorted probability maps provide a visualization of the systematic uncertainty affecting each \ac{pdf}.
    }\label{fig:background-model}
\end{figure*}

The \ac{lar} veto model is applied to each background contribution before analysis cuts, as established by the background model. 
The distribution of \al events originating from the \ac{hpge} readout contact is negligibly affected by the \ac{lar} veto cut, as the \al particles are expected to originate from the germanium surfaces themselves, where the light collection is poor and any remaining light output from the recoiling nucleus would be quenched~\cite{Doke2002}. 
Thus, the \al model has been imported as-is from~\cite{Agostini2020}. 
All predictions after the \ac{lar} veto cut are corrected for accidental rejection due to random coincidences of \SI{2.7}{\percent}~\cite{Agostini2020a}.
As in~\cite{Agostini2020}, a Poisson likelihood is used to compare the detector-type specific energy spectra of single-detector events that survive the \ac{lar} veto cut, corresponding to an exposure of \SI{61.4}{\kg\year} of \ac{gerda} \ptwo data\footnote{%
Additional \SI{1.2}{\kg\year} of data from the last physics run of \ptwo, neglected in previous works~\cite{Agostini2020}, is considered here.}, with a linear combination of background \acp{pdf}.
Statistical inference is carried out to determine the coefficients of the admixture that best describe the data. 
In a Bayesian setting, posterior probability distributions of background source intensities resulting from the model before cuts are fed as prior information\footnote{%
The procedure is not statistically rigorous, since data before and after the cut are not fully independent. Nevertheless, the methodology is considered acceptable, as the intention is to carry out a qualitative comparison.}, with the exception of \nuc{K}{42}. 
The distribution of \nuc{K}{42} ions in \ac{lar} is knowingly inhomogeneous, as ion drifts are induced by electric fields (generated by high-voltage cables and detectors) and convection. 
The spatial distribution is at present unknown.
As a matter of fact, rough approximations have been adopted to describe it in the background model~\cite{Agostini2020}. 
Given the inhomogeneity of the \ac{lar} veto response itself, a significant mismatch of the predicted event suppression between simulation and data is expected. Hence, we adopt uninformative, uniform priors for the \nuc{K}{42} source intensities.
%For the same reason, at second order, the event suppression mis-modeling is also expected to be energy dependent. 
The Monte Carlo Markov Chains are run with the BAT software~\cite{BATv1} to compute posterior distributions and build knowledge update plots.

Substantial agreement between event suppression predicted by the \ac{lar} veto model and data is found: the posterior distributions are compatible with the priors, where non-uniform, at the 1--2\;$\sigma$ level.
As anticipated, the \nuc{K}{42} activity differs significantly from the data before the \ac{lar} veto cut. 
%We stress out that, as a consequence of the known systematic uncertainties affecting the modeling of this background source, the extracted activity cannot be interpreted as the actual \nuc{K}{42} concentration in the \ac{gerda} \ac{lar}.
Based on this background decomposition, the event survival probability after \ac{lar} veto predicted in the \ac{onbb} decay analysis window\footnote{%
The \ac{onbb} decay analysis window is defined as the energy window from \SI{1930}{\keV} to \SI{2190}{\keV}, excluding the region around \qbb ($\qbb \pm 5$\;\SI{}{\keV}) and the intervals \SI{2104 \pm 5}{\keV} and $2119 \pm 5$\;\SI{}{\keV}, which correspond to known \ga lines from \nuc{Tl}{208} and \nuc{Bi}{214}.} 
is about \SI{0.3}{\percent} for \nuc{Th}{228}, \SI{15}{\percent} for \nuc{U}{238} and \SI{10}{\percent} for \nuc{Co}{60}. 
As a final remark, we stress that the quoted event suppression in the \ac{onbb} decay region can be affected by background modeling uncertainties (\eg source location, surface-to-volume activity or the exact \nuc{K}{42} spatial distribution) whose evaluation is out of the scope of this work. 
As such, they must be taken \emph{cum grano salis} and can not be generalized for different experimental conditions.

In the energy region dominated by the \ac{tnbb} decay, \ie from the \nuc{Ar}{39} endpoint at \SI{565}{\keV} to the \ac{bb} Q-value \SI{2039}{\keV}, excluding the intense but narrow potassium peaks, the ratio between the number of \ac{tnbb} events and the residual background influences the sensitivity of searches for \ac{bb} exotic decay modes~\cite{GERDA:2022ffe}.
The signal-to-background ratio is about 2 in data before analysis cuts~\cite{Agostini2020} and improves to about 18 after applying the \ac{lar} veto cut.

The obtained background decomposition is displayed in \myref{fig:background-model}. 
Bands constructed by using maximally distorted maps (\ie distortion parameter $\alpha = 0.5$ and 1.5), obtained with the procedure described in \myref{sec:distortion-studies}, are displayed for every background contribution to represent the systematic uncertainty. 
The ratio between data and best-fit model normalized by the expected statistical fluctuation in each bin, is shown below in the bottom panel.
No significant deviations are observed, beyond the expected statistical fluctuations.

The data shown in \myref{fig:background-model} is available in ASCII format as Supplemental Material~\cite{Fig7Data}.

\section{Conclusions}\label{sec:conclusions}

This paper describes the methodology, optimization and application of the light collection model as developed for the \ac{gerda} \ac{lar} scintillation light read-out.
It is based on an \emph{ansatz} that decouples the light propagation from non-optical simulations, using photon detection probability maps.
The model has been optimized using low-activity \nuc{Th}{228} calibration data. 
It allows predictions of the \ac{lar} veto event rejection, which is central for analyses of the \ac{tnbb} spectrum~\cite{GERDA:2022ffe}, including a precise determination of the \nuc{Ge}{76} half-life as well as a search for new physics phenomena.
Even though detailed insight into the heterogeneous setup was granted, one short-coming seems imminent: 
certain parts of the probability maps are extrapolated, as they remain largely unprobed by the available calibration data.
The systematic uncertainty associated with this problem has been evaluated using analytical distortions of the veto response.
It is left to the upcoming \ac{legend} experiment to reduce this uncertainty by dedicated calibration measurements, that will elevate their \ac{lar} instrumentation from a binary light/no-light veto to a full-fledged detector.

\begin{acknowledgement}
    The \ac{gerda} experiment is supported financially by
        the German Federal Ministry for Education and Research (BMBF),
        the German Research Foundation (DFG),
        the Italian Istituto Nazionale di Fisica Nucleare (INFN),
        the Max Planck Society (MPG),
        the Polish National Science Centre (NCN),
        the Foundation for Polish Science (TEAM/2016-2/17),
        the Russian Foundation for Basic Research,
        and the Swiss National Science Foundation (SNF).
    This project has received funding/support from the European Union's
    \textsc{Horizon 2020} research and innovation programme under
    the Marie Sklodowska-Curie grant agreements No 690575 and No 674896.
    This work was supported by the Science and Technology Facilities Council, part
    of the U.K. Research and Innovation (Grant No. ST/T004169/1).
    The institutions acknowledge also internal financial support.
    
    The \ac{gerda} collaboration thanks the directors and the staff of LNGS for their continuous strong support of the \ac{gerda} experiment.
\end{acknowledgement}

\begingroup
\setlength{\emergencystretch}{2em}
\printbibliography
\endgroup

\begin{figure*}
    \centering
    \includegraphics[width=\textwidth]{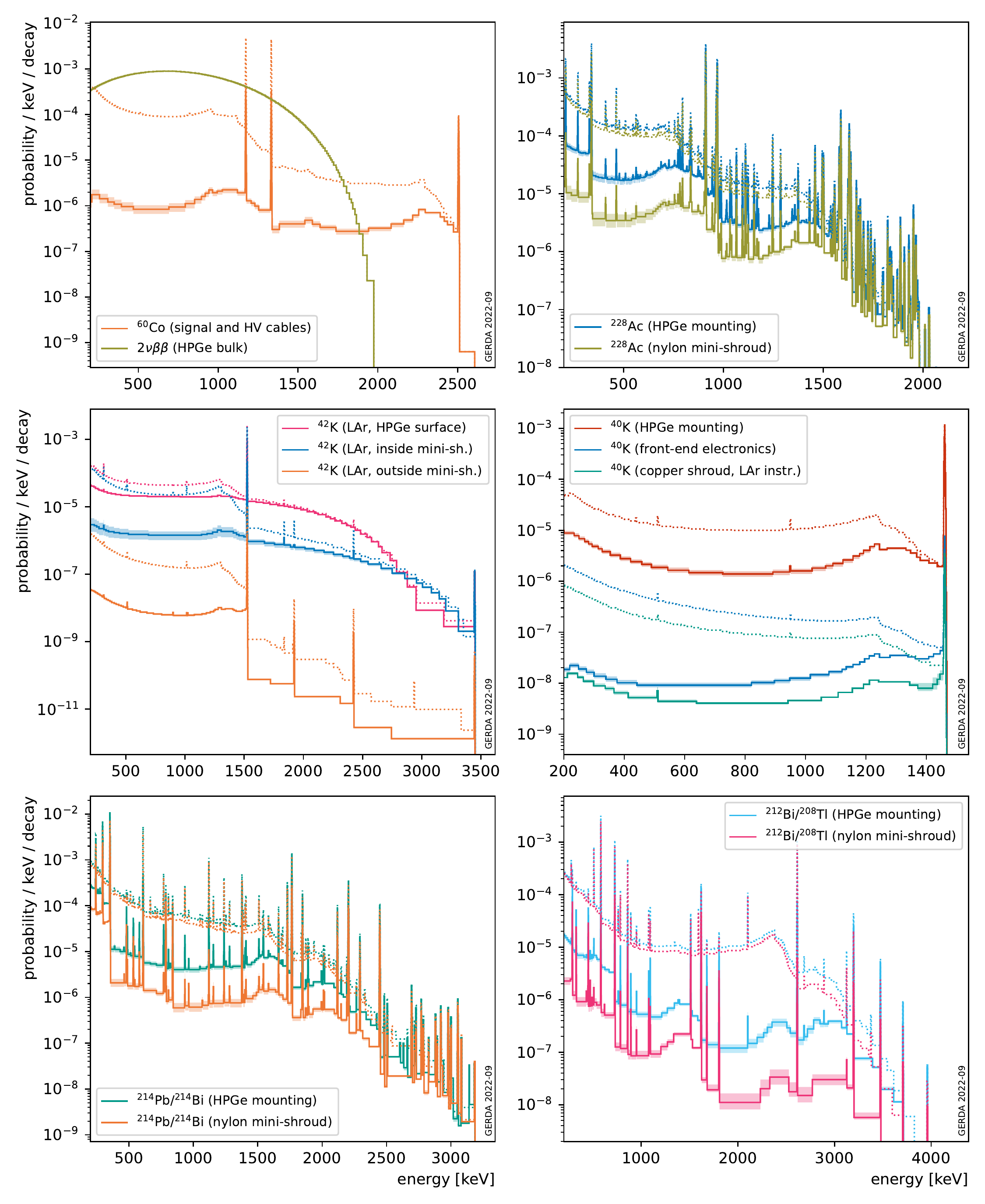}
    \caption{
        Probability density functions (\acp{pdf}, normalized to the number of simulated primary decays) for a representative selection of background and signal event sources in the \ac{gerda} \ptwo setup as detected by \ac{hpge} detectors and surviving the \ac{lar} veto cut, as predicted by the model presented in this document.
        Model uncertainties are shown as bands of lighter color.
        \Acp{pdf} before the cut~\cite{Agostini2020} (dotted lines) are overlaid for comparison.
        The reader is referred to~\cite{Agostini2020} (Figure 1) for a detailed documentation of the simulated setup.
        A variable binning is adopted for visualization purposes.
    }\label{fig:pdfs}
\end{figure*}

\end{document}

%% file: abbreviations.tex
\usepackage{acro}
\usepackage{xspace}
\usepackage{xstring}
\usepackage{isomath}
\usepackage{alphabeta}

\DeclareAcronym{bb}    {short = \ensuremath{\textnormal{\beta\beta}},     long = double beta}
\DeclareAcronym{onbb}  {short = \ensuremath{0\textnormal{\nu\beta\beta}}, long = neutrinoless double beta}
\DeclareAcronym{tnbb}  {short = \ensuremath{2\textnormal{\nu\beta\beta}}, long = two neutrino double beta}
\DeclareAcronym{roi}   {short = ROI, long = region of interest}
\DeclareAcronym{gerda} {short = \scshape{Gerda},  long = Germanium Detector Array}
\DeclareAcronym{legend}{short = LEGEND, long = Large Enriched Germanium Experiment for Neutrinoless double beta Decay, tag = experiment}
\DeclareAcronym{lngs}  {short = LNGS,             long = Laboratori Nazionali del Gran Sasso}
\DeclareAcronym{llama} {short = LLAMA,            long = \acs*{legend} Liquid  Argon  Monitoring  Apparatus}
\DeclareAcronym{hpge}  {short = HPGe, long = high-purity germanium}
\DeclareAcronym{sipm}  {short = SiPM, long = silicon photomultiplier}
\DeclareAcronym{pmt}   {short = PMT,  long = photomultiplier tube}
\DeclareAcronym{tpb}   {short = TPB,  long = tetraphenyl butadiene}
\DeclareAcronym{ptfe}  {short = PTFE, long = polytetrafluoroethylene}
\DeclareAcronym{wls}   {short = WLS,  long = wavelength-shifting}
\DeclareAcronym{pde}   {short = PDE,  long = photon detection efficiency}
\DeclareAcronym{pe}    {short = p.e., long = photo-electron, short-plural = }
\DeclareAcronym{psd}   {short = PSD,  long = pulse shape discrimination}
\DeclareAcronym{vuv}   {short = VUV,  long = vacuum-ultraviolet}
\DeclareAcronym{lar}   {short = LAr,  long = liquid argon}
\DeclareAcronym{cl}    {short = C.L.,           long  = confidence level}
\DeclareAcronym{pdf}   {short = \itshape{pdf},  long  = probability density function}
\DeclareAcronym{pmf}   {short = \itshape{pmf},  long  = probability mass function}
\DeclareAcronym{geant} {short = \scshape{Geant}4, long  = GEometry ANd Tracking}
\DeclareAcronym{mage}  {short = MaGe,             long  = Majorana-Gerda}
\DeclareAcronym{fep}   {short = FEP,              long  = full energy peak}
\DeclareAcronym{dep}   {short = DEP,              long  = double escape peak}
\DeclareSIUnit\year{yr}
\DeclareSIUnit\pe{p.e.}

\NewDocumentCommand{\larmap}    {}      {\ensuremath{\xi(\vec{x})}\xspace}
\NewDocumentCommand{\be}        {}      {\ensuremath{\textnormal{\beta}}\xspace}
\NewDocumentCommand{\ga}        {}      {\ensuremath{\textnormal{\gamma}}\xspace}
\NewDocumentCommand{\al}        {}      {\ensuremath{\textnormal{\alpha}}\xspace}
\NewDocumentCommand{\pone}      {}      {{Phase\,I}\xspace}
\NewDocumentCommand{\ptwo}      {}      {{Phase\,II}\xspace}
\NewDocumentCommand{\qbb}       {}      {\ensuremath{Q_\textnormal{\beta\beta}}\xspace}
\NewDocumentCommand{\nuc}       {m m o} {\ensuremath{^{#2}\IfNoValueTF{#3}{}{_{#3}}\textrm{#1}}\xspace} 
\NewDocumentCommand{\hairsp}    {}      {\hspace{1pt}} 
\NewDocumentCommand{\waitforit} {}      {\textendash\xspace} 
\NewDocumentCommand{\ie}        {}      {\mbox{\textit{i.\hairsp{}e.}}\xspace}
\NewDocumentCommand{\eg}        {}      {\mbox{\textit{e.\hairsp{}g.}}\xspace}

\NewDocumentCommand{\apriori}   {}      {\mbox{\textit{a priori}}\xspace}
\NewDocumentCommand{\Apriori}   {}      {\mbox{\textit{A priori}}\xspace}
\NewDocumentCommand{\adhoc}     {}      {\mbox{\textit{ad-hoc}}\xspace}

\NewDocumentCommand{\orderSI}   {m m}   {\SI[parse-numbers = false]{\mathcal{O}(#1)}{#2}}
\NewDocumentCommand{\noparseSI} {m m}   {\SI[parse-numbers = false]{#1}{#2}}
\NewDocumentCommand{\myref}{m}{%
  \IfSubStr{#1}{fig:}{Fig.~\ref{#1}}{}%
  \IfSubStr{#1}{chap:}{Chap.~\ref{#1}}{}%
  \IfSubStr{#1}{sec:}{Sec.~\ref{#1}}{}%
  \IfSubStr{#1}{tab:}{Tab.~\ref{#1}}{}%
  \IfSubStr{#1}{eq:}{Eq.~\ref{#1}}{}%
  \IfSubStr{#1}{app:}{App.~\ref{#1}}{}%
}

\NewDocumentCommand{\acsu}  {m} {\acs{#1}\acuse{#1}\xspace}
\NewDocumentCommand{\acfu}  {m} {\acf{#1}\acuse{#1}\xspace}
\NewDocumentCommand{\tetratex}  {} {Tetratex\textsuperscript{\textregistered}\xspace}

\DTMsavetimestamp{startptwo}{2015-12-20T14:42:53Z}  %1450622573 in file gerda-run0053-20151220T144252Z
\DTMsavetimestamp{endptwo}{2018-04-16T07:35:18Z}    %1523864118 in file gerda-run0093-20180416T043538Z
\DTMsavetimestamp{startptwop}{2018-07-18T14:50:34Z} %1531925434 in file gerda-run0095-20180718T145030Z
\DTMsavetimestamp{endptwop}{2019-11-11T22:46:45Z}   %1573512405 in file gerda-run0114-20191111T200216Z
\DTMsavetimestamp{pca68}{2016-07-11T08:56:29Z}      %1468227389 in file gerda-run0068-20160711T085629Z
\DTMsavetimestamp{pca76}{2017-02-05T12:55:28Z}      %1486299328 in file gerda-run0076-20170205T125528Z

\usepackage{xcolor}

%% file: authors.tex
\author{%
The \mbox{\protect{\sc{Gerda}}} collaboration\thanksref{corrauthor} \and \\[4mm]
M.~Agostini\thanksref{UCL} \and
A.~Alexander\thanksref{UCL} \and
G.R.~Araujo\thanksref{UZH} \and
A.M.~Bakalyarov\thanksref{KU} \and
M.~Balata\thanksref{ALNGS} \and
I.~Barabanov\thanksref{INRM} \and
L.~Baudis\thanksref{UZH} \and
C.~Bauer\thanksref{HD} \and
S.~Belogurov\thanksref{ITEP,INRM,alsoMEPHI} \and
A.~Bettini\thanksref{PDUNI,PDINFN}
% ...
}

\authorrunning{the \textsc{Gerda} collaboration}
\thankstext{corrauthor}{\emph{correspondence:} \texttt{gerda-eb@mpi-hd.mpg.de}}
\thankstext{alsoMEPHI}{\emph{also at:} NRNU MEPhI, Moscow, Russia}
\thankstext{nowDuke}{\emph{present address:} Duke University, Durham, NC USA}
\thankstext{alsoLev}{\emph{also at:} Moscow Inst. of Physics and Technology, Russia}
\thankstext{nowIKZ}{\emph{present address:} Leibniz-Institut f{\"u}r Kristallz{\"u}chtung, Berlin, Germany}
\thankstext{alsoDubna}{\emph{also at:} Dubna State University, Dubna, Russia}
\thankstext{nowLehnert}{\emph{present address:} Nuclear Science Division, Berkeley, USA}
\thankstext{nowKermaidic}{\emph{present address:} Université Paris-Saclay, CNRS/IN2P3, IJCLab, 91405
Orsay, France}

\institute{%
INFN Laboratori Nazionali del Gran Sasso, Assergi, Italy\label{ALNGS}
% ...
\and
Max-Planck-Institut f{\"u}r Kernphysik, Heidelberg, Germany\label{HD} \and
Department of Physics and Astronomy, University College London, London, UK\label{UCL}
% ...
\and
Institute for Nuclear Research of the Russian Academy of Sciences, Moscow, Russia\label{INRM} \and
Institute for Theoretical and Experimental Physics, NRC ``Kurchatov Institute'', Moscow, Russia\label{ITEP} \and
National Research Centre ``Kurchatov Institute'', Moscow, Russia\label{KU}
% ...
\and
Dipartimento di Fisica e Astronomia, Universit{\`a} degli Studi di 
Padova, Padua, Italy\label{PDUNI} \and
INFN  Padova, Padua, Italy\label{PDINFN}
% ...
\and
Physik-Institut, Universit{\"a}t Z{\"u}rich, Z{u}rich, Switzerland\label{UZH}
}